\documentclass{elsart}
\usepackage{graphics}
\usepackage{epsfig}

\begin{document}

\begin{frontmatter}



\title{Mathematical Model of Easter Island Society Collapse}


\author{ M. Bologna  }
\author{  J. C. Flores   }

\address{Instituto de Alta Investigaci\'{o}n, Universidad de
  Tarapac\'{a}-Casilla 7-D Arica, Chile}

\begin{abstract}
In this paper we consider a mathematical model for the evolution
and collapse of the Easter Island society, starting from the fifth
century  until the last period of the society collapse (fifteen
century). Based on historical reports, the available primary
sources consisted almost exclusively on the trees. We describe the
inhabitants and the resources as an isolated system and both
considered as dynamic variables. A mathematical analysis about why
the structure of the Easter Island community collapse is
performed. In particular, we analyze the critical values of the
fundamental parameters driving the interaction humans-environment
and consequently leading to the collapse. The technological
parameter, quantifying the exploitation of the resources, is
calculated and applied to the case of other extinguished
civilization (Cop\'an Maya) confirming, with a sufficiently
precise estimation, the consistency of the adopted model.
\end{abstract}

\end{frontmatter}


\section{Introduction}

\bigskip

\bigskip

\label{1intr1} Easter Island history is a very famous example of an evolved
human society that collapsed for over exploiting its fundamental resources
\cite{brander,dalton,anderies} that in this case were essentially in palm
trees. It were covering the island\cite{flenley} when, few dozens of
individuals, first landed around 400 A.D. Its advanced culture was developed
in a period of one thousand years approximately. Its ceremonial rituals and
associated construction were demanding more and more natural resources
especially palm trees. The over exploitation of this kind of tree, very
necessary as a primary resource (tools construction, cooking, erosion
barrier, etc.) was related with the collapse. In this paper, a mathematical
model concerning growing and collapse of this society is presented.
Different to usual like Lotka-Volterra models\cite{has,mur} where the
carrying capacity variation becomes from external natural forces, in this
work it is directly connected with the population dynamics. Namely,
population and carrying capacity are interacting dynamics variables, so
generalizing the Leslie model prey-predator.

The general mathematical treatment of a model describing such a
complex society is a very hard task and probably not unique. Our
aim is to settle the most simple model describing with acceptable
precision the evolution of Easterner society. With the idea of
writing a model that could be generalized to a more complex
system, we first divide the elements into two categories: the
\textit{resource} quantity $R_{i}$ with $i=1,2\cdots k$ and the
\textit{inhabitants} numbers (species) $N_{i}$ with $i=1,2\cdots
m$. With the concept of resources we are meaning resources in a
very large sense, it could be oil, trees, food and so on. The
several kind of resources are described by the index $i$. In
similar way, with the concept of inhabitants, we are meaning
different species of animals or internal subdivision of human
being in country or town or even tribes. Leaving the idea of a
constant quantity of resources that leads to the logistic equation
for the number on inhabitants\cite{logistic}, the aim of this
paper is to include in the dynamical description of the time
evolution of the system the resources too which can not be
considered as constant. A generalization of the logistic equation
to an arbitrary number of homogeneous species interacting among
the individuals (with non constant resources) can be written as

\begin{equation}
\frac{d}{dt}N_{i}=r_{i}N_{i}\left[ 1-\frac{N_{i}}{N_{ci}\left( R_{1}\cdots
R_{k}\right) }\right] -\sum\limits_{j=1\,,i\neq j}^{m}\chi _{ij}N_{j}N_{i}.
\label{log-inter}
\end{equation}%
Where $r_{i}$ is the usual growing rate for species $i$. In the denominator
it appears the carrying capacity of the system with respect to the number of
inhabitants $N_{ci}\left( R_{1}\cdots R_{k}\right) $. Beside the dependence
on the resources $R_{i}$ we could have also a dependence on an other species
that would be then a \textquotedblright resource\textquotedblright\ for some
other species. This fact is expressed even by the quantities $\chi _{ij}$
that in general are not symmetric expression ($\chi _{ij}\neq \chi _{ji}$)
since that the prey is a resource for the predator and not the inverse.
Similarly, for the resources we have:

\begin{equation}
\frac{d}{dt}R_{i}=r_{i}^{\prime }R_{i}\left[ 1-\frac{R_{i}}{R_{ci}}\right]
-\sum\limits_{j=1}^{m}\alpha _{ij}N_{j}R_{i}.  \label{res-inter}
\end{equation}%
It is clear that set of equations (\ref{res-inter}) could be formally
included into Eq. (\ref{log-inter}) redefining the quantities $\alpha _{ij}$%
. Nevertheless, we shall keep this distinction for the sake of
clarity especially referred to resources, such as trees, oil or
oxygen, where the carrying capacity is not determined by other
species and can be considered a constant ($R_{ci}$). Also the
meaning of the parameters $r_{i}^{\prime }$ is more or less the
same of the analogous parameters $r_{i}$. It is suitable to define
$r_{i}^{\prime }$ as \textquotedblright renewability
ratios\textquotedblright , since describe the capacity of the
resources to renew itself and clearly are depending on the kind of
resource. For example, the renewability of the oil is clearly zero
since the period of time to get oil from a natural process is of
the order of geological time-scale processes. In general all
parameters of Eqs. (\ref{log-inter}) and (\ref{res-inter}) are
time dependent, including stochasticity. We can assume reasonably
slowly time-varying for the ancient societies so that we can
consider it as constant\cite{flores}, particularly the $\alpha
$'s. Anyway the set of $\alpha _{ij}$ is worthy of a more detailed
discussion. We can call this set of parameters
\textit{technological parameters} in the sense that they carry the
information about the capacity to exploit the resources of the
habitat. We shall see in the next section that the technological
parameters combined with the renewability ratios will be the key
point to decide wether, or not, a society is destined to collapse.

\section{Easter island collapse model}

\label{2east2} The particular history of Easter Island society presents
several advantages for modelling its evolution \cite{bologna}. In fact it
can be with very good approximation considered a closed system. The peculiar
style of life and culture allows us to consider a basic model where trees
are essentially the only kind of resources. Many of activities of the
ancient inhabitants involved the trees, from building and transport the
enormous Moai, to build boats for fishing, etc. In fact the cold water was
not adapt to the fish life and the impervious shape of the coast made
difficult fishing. Finally from historical reports it can be inferred that
the inhabitants did not change the way to exploit their main resource, even
very near to exhaust it so that we can consider the technological parameter
as constant. Considering Eqs. (\ref{log-inter}) and (\ref{res-inter}) for
one inhabitant species and one kind of resource, we obtain:

\begin{equation}
\frac{d}{dt}N=rN\left[ 1-\frac{N}{N_{c}\left( R\right) }\right] ,
\label{log-inter1}
\end{equation}

\begin{equation}
\frac{d}{dt}R=r^{\prime }R\left[ 1-\frac{R}{R_{c}}\right] -\alpha NR,
\label{res-inter1}
\end{equation}%
where we introduce the notation: $\alpha _{11}\equiv \alpha $. The unknown
function $N_{c}\left( R\right) $ has to satisfy few properties. For a
quantity of primary unlimited resource, $R\rightarrow \infty $, even $%
N_{c}\left( R\right) \rightarrow \infty $ that means that the
population can grow unlimited too. In the opposite case
$R\rightarrow 0$ clearly also the population must vanish,
$N_{c}\left( R\right) \rightarrow 0$, and finally when the
resource is constant we are back to the ordinary equation
(logistic) so that $N_{c}\left( R\right) = \textrm{ const}$. It is
clear that the choice of this relation is quite arbitrary but
following the simplicity criteria we can select $N_{c}\left(
R\right) =\beta R$, where $\beta $ is a positive parameter. This
choice formalizes the intuition that the maximum number of
individuals tolerated by a niche is proportional to the quantity
of resources. We note that in (\ref{res-inter1}) the interacting
term depend
on the variable $R$. Namely, for $R=0$ no variation of resource exist ($%
\frac{d}{dt}R=0$) corresponding to a biological criterion and different from
this one of reference\cite{basener}. A more sophisticate model should
include also fishing as resource and consider for $N_{c}\left( R\right) $ an
expression such as $N_{c}\left( R\right) =\beta R+R_{f}$ where $R_{f}$ is
the fishing carrying capacity. This resource was limited near the coast and
could not be fully exploited without boats, so that, we are going to neglect
this resource. We can rewrite Eqs. (\ref{log-inter1}) and (\ref{res-inter1})
as:

\begin{equation}
\frac{d}{dt}N=rN\left[ 1-\frac{N}{\beta R}\right] ,  \label{log-inter2}
\end{equation}

\begin{equation}
\frac{d}{dt}R=r^{\prime }R\left[ 1-\frac{R}{R_{c}}-\alpha
_{E}N\right] ,\,\,\,\,\textrm{where }\,\,\alpha _{E}\equiv
\frac{\alpha }{r^{\prime }}. \label{res-inter2}
\end{equation}%
The dimensionless parameter $\alpha _{E}$ is the ratio between the
technological parameter $\alpha $, representing the capability of to exploit
the resources, and $r^{\prime }$ the renewability parameter representing the
capability of the resources to regenerate. We will call $\alpha _{E}$
deforestation parameter since it gives a measure of the rapidity with which
the resources are going to exhaust and then a measure of the reversibility
or irreversibility of the collapse. Using the historical data we can have an
estimation of the parameters. At the origin ($t=0$) we can assume that the
trees were covering the entire island surfaces of 160 km$^{2}$. When the
first humans arrived to the island, around the 400 A.D., their number was of
the order of few dozens of individuals and it grew until to reach the
maximum $N_{M}\sim 10000$ around the 1300 A.D.

Finding the equilibrium points of Eqs. (\ref{log-inter2}) and (\ref%
{res-inter2}) we obtain (see section III, for stability):

\begin{equation}
N_{0}=0,\,\,\,\,R_{0}=R_{c}\,\,\,\,(\textrm{unstable-saddle-point})
\label{equin}
\end{equation}

\begin{equation}
N_{e}=\frac{\beta R_{c}}{1+\alpha _{E}\beta R_{c}},\,\,\,\,R_{e}=\frac{R_{c}%
}{1+\alpha _{E}\beta R_{c}}\,\,\,\,(\textrm{stable}).
\label{equir}
\end{equation}%
While the point $(N_{0},R_{0})$ of Eq. (\ref{equin}) represents the trivial
fact that in absence of human being the number of trees is constant
(carrying capacity). Eq. (\ref{equir}) describes the fact that, due to the
interaction humans-environment, the more interesting equilibrium point $%
(N_{e},R_{e})$ does not coincide with $(N_{c},R_{c})$ with $N_{c}=\beta
R_{c} $ since $\alpha \neq 0$. To study the stability of the point $%
(N_{e},R_{e})$, we have linearize the system of Eqs. (\ref{log-inter2}) and (%
\ref{res-inter2})around the equilibrium point. In fact, in the next section
we shall show that it is a stable equilibrium point.

\section{Equilibrium points and stability}

\label{equilibrium}

Let us first cast Eqs. (\ref{log-inter2}) and (\ref{res-inter2})
in term of dimensionless quantities; setting $\nu (t)=N/N_{c}$,
$\varrho (t)=R/R_{c}$ and $\tau =rt$, we have:

\begin{equation}
\frac{d}{d\tau }\nu =\nu \left[ 1-\frac{\nu }{\varrho }\right]  \label{adnu}
\end{equation}

\begin{equation}
\frac{d}{d\tau }\varrho =\bar{r}\varrho \left[ 1-\varrho -\bar{\alpha}\nu %
\right] ,\,\,\,\,\bar{\alpha}\equiv \alpha _{E}N_{c},\,\,\,\,\bar{r}\equiv
\frac{r^{\prime }}{r}.  \label{adro}
\end{equation}
For sake of clarity we rewrite also the equilibrium point:
\begin{equation}
\nu _{e}=\frac{1}{1+\bar{\alpha}},\,\,\,\,\varrho _{e}=\frac{1}{1+\bar{\alpha%
}}  \label{equirad}
\end{equation}
with obvious meaning of the symbols. Perturbing the equilibrium point $\nu
(\tau )=\nu _{e}\left[ 1+\eta (\tau )\right] $ and $\varrho (\tau )=\varrho
_{e}\left[ 1+\varepsilon (\tau )\right] $ with $\eta (\tau )$ and $%
\varepsilon (\tau )$ infinitesimal functions, after straightforward algebra
we obtain Eq. (\ref{equir}):

\begin{equation}
\frac{d}{d\tau }\eta =-\eta +\varepsilon  \label{lineta1}
\end{equation}

\begin{equation}
\frac{d}{d\tau }\varepsilon =-\frac{\bar{r}}{1+\bar{\alpha}}\left[ \bar{%
\alpha}\eta +\varepsilon \right]  \label{lineps1}
\end{equation}
The eigenvalues of the system are:

\begin{equation}  \label{eig}
\lambda_{1,2}=\frac{-(1+\bar{\alpha}+ \bar{r})\pm\sqrt{\left(1+\bar{\alpha}+%
\bar{r}\right)^{2} -4\bar{r}(1+\bar{\alpha})^{2}}}{2(1+\bar{\alpha})}
\end{equation}
Restricting ourself to the case of positive values of $\bar{\alpha}$, Eq. (%
\ref{eig}) shows that both the eigenvalues always have a real negative part,
so that the equilibrium point $(\nu_e,\varrho_e)$ is a stable equilibrium
point. More in detail we have that for

\[
\bar{r}\leq \frac{1}{4},\,\,\bar{\alpha}\geq 0\,\,\,\,\textrm{and for}\,\,\,\,%
\bar{r}>\frac{1}{4},\,\,\bar{\alpha}\leq \frac{\left( \sqrt{\bar{r}}%
-1\right) ^{2}}{2\sqrt{\bar{r}}-1}
\]
the eigenvalues are real and negative so that the equilibrium point is
reached in exponential damped way, otherwise the eigenvalues acquire an
imaginary part and the system reach the equilibrium point via exponential
damped oscillations.

\begin{figure}[h]
\par
\begin{center}
\epsfig{file=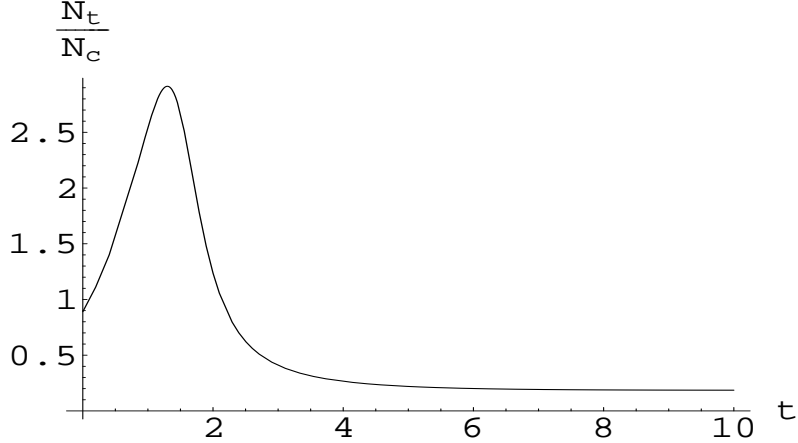,height=6cm,width=12cm,angle=0}
\end{center}
\caption{\label{figlor} $N(t)/N_c$ versus time. In this numerical
example $N(0)/N_c\sim 0.8$ and $N_{Max}/N_c\sim 3$}
\end{figure}

\section{Collapse condition}

Even if, mathematically speaking, the stable point $(N_{e},R_{e})$
is an acceptable result we have to take in account the biological
constraints that allow to a specie to survive. A reasonable number
of individuals is required for viability of a given species
\cite{beg,bel}. This is so because genetic diversity, social
structures, encounters, etc., need a minimum numbers of
individuals since under this critical numbers the species is not
viable and collapse. It is worthy to stress that while the trees
can reach an equilibrium point without the humans, Eq.
(\ref{equin}), the opposite does not hold, as stated by Eq.
(\ref{equir}).

Calling the minimum number of humans $N_{min}$ we can find a upper bound for
the parameter $\alpha _{E}$ so that a civilization can survive. Imposing the
condition that at the equilibrium point $N_{e}\geq N_{min}$ we obtain:

\begin{equation}
\alpha _{E}\geq \frac{1}{N_{min}}-\frac{1}{\beta R_{c}}\textrm{
(collapse condition)}.  \label{bond}
\end{equation}%
As further simplification of inequality (\ref{bond}), we assume that $%
N_{min}\ll \beta R_{c}$ and we find that $\alpha _{E}N_{min}\geq 1$ or $%
\alpha N_{min}\geq r^{\prime }$. It is a natural condition since it tells
that collapse exists when the production rate $r^{\prime }$ is minor than
the deforestation rate $\alpha N_{min}$.

More in general it can be showed, numerically, that considering
the standard case with the starting population number
$N(0)<N_{c}=\beta R_{c}$, we can have a solution that can exceed
the value $\beta R_{c}$ (depending on initial conditions $N(0)$
and $\frac{dN}{dt}|_{t=0}$). It is worthy to stress that in the
case of ordinary logistic map the population number never can
exceed this limit value\cite{logistic}. In the example of Fig.
\ref{figlor}, the maximum of $N$ is reached at a value that is
almost three times $\beta R_{c}$. Then, according to the region of
the parameters that we are considering, the paths to the final
equilibrium is exponentially fast, reaching eventually the point
$N_{min}$ and collapsing.

\section{Deforestation rate estimation}

As we saw in the previous section, the collapse condition (15) gives a
sufficient condition on the deforestation rate per individual $\alpha$ . On
other hand, the last period of tree extinction was governed essentially by
the deforestation rate. In this way, we have the rate of tree extinction

\begin{equation}
\frac{1}{R}\frac{dR}{dt}\sim -\alpha N.  \label{16}
\end{equation}%
As discussed, the path to the equilibrium point is exponentially fast, so
that a rough estimation of the left side of Eq. (\ref{16}) is the time scale
of the deforestation, $\tau _{F}$, while the right side can be taken at the
end of the collapse process (the equilibrium point):
\begin{equation}
\frac{1}{\tau _{F}}\sim \alpha N_{F},  \label{17}
\end{equation}%
being $N_{F}$ the final number of individuals. It can be deduced \cite%
{pointing} that the range of is $\tau _{F}\sim 100$ yrs. to $\tau _{F}\sim
300$ yrs. and $N_{F}\sim 3000$, the rate of deforestation (per individual)
could be estimated as:
\begin{equation}
\alpha \sim \frac{1}{\tau _{F}N_{F}}\textrm{ ( yrs.
individual)}^{-1}
\end{equation}%
giving a range
\begin{equation}
1.1\times 10^{-6}<\alpha <3.3\times 10^{-6}\textrm{ ( yrs.
individual)}^{-1}.
\end{equation}%
This estimation has validity in the case of exponential decay which is the
our case, as it has been showed by the analysis performed in Sec. \ref%
{equilibrium}. Assuming the number of trees as proportional to the area $A$
we can now estimate the rate-deforestation-area. The island has a surface at
order of $A\sim 160$ km$^{2}$ and initially it can be supposed that was
covered of trees so that we can estimate:
\begin{equation}
0.5<\frac{dA}{dt}\sim \frac{A_{0}}{\tau _{F}}<1.6\,\,(\textrm{km}^{2}/\textrm{yrs%
}).
\end{equation}%
As comparison we can consider that in the last 500 years the
deforestation of amazonian forest rate is 15
$(\textrm{km}^{2}/\textrm{yrs})$. Considering that in 500 years
the deforestation technology became more and more efficient,
especially in the last century, we can consider it as an upper
limit, giving us an idea of the technological change. In the human
history there are several examples of over exploiting the natural
resources even if not so known as Easter island. In particular, the Cop\'{a}%
n Maya history\cite{wil} has certain similarity with respect to the
technology level and the over exploiting of the natural resources. In short,
this ancient civilization reached almost 20.000 individuals and declined to
5000 individuals in the 9th century. Using the estimation of the
technological parameter $\alpha $ obtained for eastern island civilization,
we get a collapse time from Eq. (\ref{17}) that is $60<\tau <180$ years. The
collapse time based on historical reports is $\tau \sim 100$, showing that
the adopted model is consistent with the available data. The estimation of
the parameter is consistent with the idea that similar civilizations, in
technological meaning, have similar capacity to exploit the natural
resources.

\section{Concluding remarks}

A mathematical model considering the interaction among carrying capacity and
population  in an isolated system has been considered. The model takes in
account the fact that a population can over exploit the carrying capacity
without saturate, a fact of relevant importance on the path leading to a
collapse of a society. Its application to the collapse of the Easter Island
civilization has been presented. An estimation of the technological
parameter $\alpha $ is obtained and applied to an other ancient
civilization, the Cop\'{a}n-Maya, with a reasonably precise expectation
about their collapse time. All confirming the consistency of the adopted
model. On the other hand, its relative reasonable prediction suggests a
possible extension to more complex system. The effort to mathematical
modelling of ancient civilizations \cite{mur,flores,hor} could be important
considering the actual human growing and resources exploitations. An
adequate equilibrium between competition, demand and exploitation is the key
of surviving.

\label{4conc4}

\section*{Acknowledgments}

The authors acknowledge support from the project UTA-Mayor 4787 (2006-2007)
and CIHDE-project.

\end{document}